\documentclass[twocolumn,showpacs,prl]{revtex4}
\usepackage{amssymb}

\usepackage{graphicx}
\usepackage{dcolumn}
\usepackage{bm}

\begin{document}

\title{Simulation and detection of Dirac fermions with cold atoms in an optical
lattice}
\author{Shi-Liang Zhu$^{1,3}$, Baigeng Wang$^{2}$, and L.-M. Duan$^{3}$}
\affiliation{$^1$ICMP and LPIT, Department of Physics, South China Normal University, Guangzhou, China\\
$^2$National Laboratory of Solid State Microstructures and
Department of Physics, Nanjing University, Nanjing, China\\
$^3$FOCUS Center and MCTP, Department of Physics, University of
Michigan, Ann Arbor, MI 48109}

\begin{abstract}
We propose an experimental scheme to simulate and observe
relativistic Dirac fermions with cold atoms in a hexagonal optical
lattice. By controlling the lattice anisotropy, one can realize
both massive and massless Dirac fermions and observe the  phase
transition between them. Through explicit calculations, we show
that both the Bragg spectroscopy and the atomic density profile in
a trap can be used to demonstrate the Dirac fermions and the
associated phase transition.
\end{abstract}

\maketitle

\ Control of ultracold atoms in an optical lattice opens up many avenues to
explore some fundamental phenomena at the forefront of condensed matter
physics\cite{ketterle,Bloch,Jaksch,Duan,Zhao}. By designing configurations of
this atomic system, one can simulate effective theories that are very
different from the microscopic atomic physics. In this paper, we add an
unusual example to the avenues of quantum simulation by showing that
ultracold atoms in an optical lattice can be used to investigate physics
associated with relativistic Dirac fermions. The ultracold atomic gas, as
the coldest setup in the universe, is one of the most non-relativistic
systems. Nevertheless, we will see that effective theories for the
quasiparticles in this system can become relativistic under certain
conditions.

We simulate Dirac fermions with single-component cold atoms in a
two-dimensional hexagonal lattice. This lattice can be formed
through interference of three laser beams, as we show below. The
physics here is closely related to the properties of electrons in
the graphene material formed with a single layer of carbon atoms
\cite{Semenoff,Novoselov,Zhang,Igor,Katsnelson,Kane,Gusynin}. The
graphene, with its emergent relativistic massless quasiparticles,
has recently raised strong interest in condensed-matter physics
\cite {Semenoff,Novoselov,Zhang,Katsnelson,Kane,Gusynin}. Compared
with the graphene, the system with cold atoms in an optical
lattice may offer more controllability. For instance, we show that
one can realize both massive and massless Dirac fermions by
controlling anisotropy of the optical lattice. This anisotropy can
be conveniently tuned through variation of the trapping laser
intensity. Under such a tuning, one can also observe a quantum
phase transition in this system. This phase transition is not
associated with any usual symmetry breaking, but instead it is
characterized by a topological change of the fermi surfaces
\cite{Lifshitz,Wen}. To detect the massive and the massless Dirac
fermions and the  phase transition between them, we calculate the
Bragg spectrum for this system as well as\ its atomic density
profile in a trap. From this calculation, we show that the
conventional atomic detection techniques based on the Bragg
spectroscopy \cite{Ketterle99} or the density profile measurement
\cite{ketterle,Zwierlein,16} can be used to demonstrate the Dirac
fermions and the  phase transition.

For cold atoms, one realizes an effectively two-dimensional system
by raising the potential barrier of the optical lattice along the
z direction to suppress the vertical tunneling between different
planes. Then, in the x-y plane, one can form a hexagonal optical
lattice with three laser beams. For instance, as shown in Ref.
\cite{Duan}, with three detuned standing-wave laser, the optical
potential is given by
\begin{equation}
V(x,y)=\sum_{j=1,2,3}V_{j}\sin ^{2}[k_{L}(x\cos \theta _{j}+y\sin \theta
_{j})+\pi /2]  \label{potential}
\end{equation}
where $\theta _{1}=\pi /3,$ $\theta _{2}=2\pi /3,$ $\theta _{3}=0$, and $%
k_{L}$ is the optical wave vector. If $V_{1}=V_{2}=V_{3}$, the
energy contour of the potential $V(x,y)$ is shown in Fig. 1(a),
where its minima (marked with the solid dots) form a standard
hexagonal lattice. It is easy to tune the potential barriers
$V_{j}$ by variation of the laser intensities along different
directions. With different $V_{j}$, one can still get a hexagonal
lattice but with a finite anisotropy. For instance, Figure 1(b)
shows the potential contour with $V_{1}=V_{2}=0.91V_{3}$, where
the neighboring sites along the horizontal direction have a larger
distance and a higher barrier. As the atomic tunneling rate in an
optical lattice is exponentially sensitive to the potential
barrier, this control provides an effective method to tune the
anisotropy of the atomic tunneling rate in this lattice.

A hexagonal lattice consists of two sub-lattices denoted by A and
B as shown in Fig. 1(c). We consider single-component fermionic
atoms (e.g., $^{40}K$,$^6Li$ etc.) in this hexagonal lattice. For
single-component fermions, the atomic collisions are negligible at
low temperature. The Hamiltonian is given by the simple form
\begin{equation}
H=-\sum_{\langle i,j\rangle }t_{ij}(a_{i}^{\dagger }b_{j}+h.c.),
\end{equation}
where $\langle i,j\rangle $ represents the neighboring sites, and $a_{i}$
and $b_{j}$ denote the fermionic mode operators for the sub-lattices A and
B, respectively. The tunneling rates $t_{ij}$ in general depend on the
tunneling directions in an anisotropic hexagonal lattice, and we denote them
as $t_{1},t_{2},t_{3}$ corresponding to the three different directions (see
Fig. 1(c)). In this work, for simplicity, we assume \ $%
t_{1}=t_{2}=t$ and $t_{3}=\beta t$, where $\beta >0$ is the
anisotropy parameter. In the following, we show that by tuning the
anisotropy $\beta $, the quasiparticles in this system change
their behaviors from massless to massive Dirac fermions, with a
quantum phase transition between the two cases.

The Hamiltonian (2) can be diagonalized with a simple extension of the
method for the graphene material \cite{Semenoff}. For the sub-lattice $A$%
, the positions of the sites can be expressed as $\mathbf{A}=m_{1}\mathbf{a}%
_{1}+m_{2}\mathbf{a}_{2}$ , where $m_{1}$ and $m_{2}$ are integers, and the
basis vectors $\mathbf{a}_{1}=(\sqrt{3},-1)(a/2)$, $\mathbf{a}_{2}=(0,a)$ ($%
a=2\pi /(\sqrt{3}k_{L})$ is the lattice spacing). The sites in the
sub-lattice $B$ are generated by a shift $\mathbf{B}=\mathbf{A}+\mathbf{b}$
with three possible shift vectors $\mathbf{b}_{1}=(1/\sqrt{3},1)(a/2)$, $%
\mathbf{b}_{2}=(1/\sqrt{3},-1)(a/2)$, and
$\mathbf{b}_{3}=(-a/\sqrt{3},0)$ (see Fig. 1(c)). The first
Brillouin zone of this system also has a hexagonal shape in the
momentum space with opposite sides identified but rotated an angle
of $\pi /6$ relative to the hexagon of the real-space lattice (see
Fig.1(d)). Corresponding to two different sites $A$ and $B$ in
each cell in real hexagonal lattice, only two of the six corners
in Fig. 1(d) are
inequivalent, usually denoted as $K$ and $K^{\prime }$. One can choose $%
\mathbf{K}=(2\pi /a)(1/\sqrt{3},1)$ and $\mathbf{K}^{\prime }=-\mathbf{K}$.
With a Fourier transform $a_{i}^{\dagger }=\frac{1}{\sqrt{N}}\sum_{\mathbf{k}%
}\exp (i\mathbf{k}\cdot \mathbf{A}_{i})a_{\mathbf{k}}^{\dagger }$ and $%
b_{j}^{\dagger }=\frac{1}{\sqrt{N}}\sum_{\mathbf{k}}\exp (i\mathbf{k}\cdot
\mathbf{B}_{j})b_{\mathbf{k}}^{\dagger }$, where $N$ is the number of sites
of the sub-lattice $A$ (or $B$), the Hamiltonian (2) simplifies to $H=\sum_{%
\mathbf{k}}[\phi (\mathbf{k})a_{\mathbf{k}}^{\dagger
}b_{\mathbf{k}}+h.c.]$,
where $\phi (\mathbf{k})=-\sum_{s=1}^{3}t_{s}\exp (i\mathbf{k}\cdot \mathbf{b%
}_{s})$. The energy eigenvalues of $H$ are given by
$E_{\mathbf{k}}=\pm |\phi (\mathbf{k})|$, which has the expression
\begin{equation}
\label{E_k} E_{\mathbf{k}}=\pm t\sqrt{2+\beta ^{2}+2\cos
(k_{y}a)+4\beta \cos \left( \sqrt{3}k_{x}a/2\right) \cos \left(
k_{y}a/2\right) }.
\end{equation}

\begin{figure}[tbph]
\label{Fig1} \includegraphics[height=7.5cm]{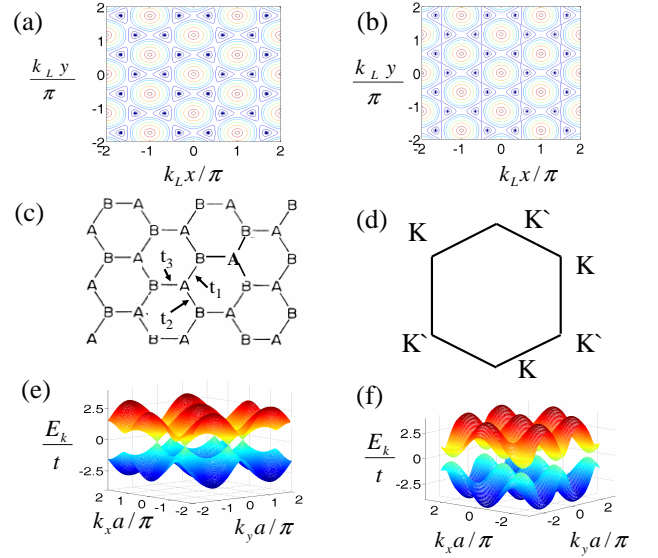} 
\caption{ (Color online) The hexagonal lattices. (a) and (b) The
contours with three potentials described in Eq.(\ref{potential}).
The minima of the potentials are denoted by the solid dots. All
$V_j^0$ are the same in (a) and $V_1^0=V_2^0=0.91 V_3^0$ in (b).
(c) Decomposition of the hexagonal lattice as two triangular
sublattices $A$ and $B$ with anisotropic tunnelling. (d) The
Brillouin zone of the hexagonal lattice. The dispersion relations
are shown in (e) for $\protect\beta=1$ (gapless state) and (f) for
$\protect\beta=2.5$ (gapped state). }
\end{figure}

The energy versus the momentum from Eq.(\ref{E_k}) is plotted in
Fig. 1(e) and 1(f) for different values of the anisotropy $\beta
$. There are two branches of curves corresponding to the \ $\pm $
sign in Eq. (3). When $0<\beta <2$, the two branches touch each
other, and around the touching points there appears a Dirac cone
structure. One has the standard cone same as the graphene
material with $\beta =1$ \cite{Semenoff,Novoselov,Zhang}, and as $\beta $ deviates from $%
1 $, the cones squeeze in $x$ or $y$ direction, but they still touch each
other. When $\beta >2$, the two branches separate with a finite energy gap $%
\Delta =\left| t\right| (\beta -2)$ between them. So, across the point $%
\beta =2$, the topology of the fermi surface changes and there is
a corresponding quantum phase transition, albeit no symmetry
breaks at this point. With this phase transition, the system
changes its behavior from a semi-metal to an insulator at the half
filling case (half filling means one atom per cell; note that each
cell has two sites). Around the half filling, the fermi surface is
close to the touching points and one can expand the
momentum $\mathbf{k}$ around one of the touching points $%
(k_{x}^{0},k_{y}^{0})$ as
$(k_{x},k_{y})=(k_{x}^{0}+q_{x},k_{y}^{0}+q_{y}).$ Up to the
second order of $q_{x}$ and $q_{y}$, the dispersion relation
(\ref{E_k}) becomes
\begin{equation}
E_{\mathbf{q}}=\pm \sqrt{\Delta
^{2}+v_{_{x}}^{2}q_{x}^{2}+v_{_{y}}^{2}q_{y}^{2}},  \label{Dispersion}
\end{equation}
where $\Delta =0$, $v_{_{x}}=\sqrt{3}\beta ta/2$ and $v_{_{y}}=ta\sqrt{%
1-\beta ^{2}/4}$ for $0<\beta <2$; and $\Delta =\left| t\right| (\beta -2)$%
, $v_{_{x}}=ta\sqrt{3\beta /2}$ and $v_{_{y}}=ta\sqrt{\beta /2-1}$ for $%
\beta >2$. This simplified dispersion relation $E_{\mathbf{q}}$ is actually
a good approximation as long as $q_{x},q_{y}\lesssim 1/2a$. We see that $E_{%
\mathbf{q}}$ represents the standard energy-momentum relation for
the relativistic Dirac particles, with $\Delta $ taking the
meaning of mass, and $v_{_{x}}$\ and $v_{_{y}}$ replacing the
light velocity. The wavefunction for the quasiparticles around the
half filling then satisfies the Dirac equation $i\hbar \partial
_{t}\Psi =H_{D}\Psi $, where the relativistic Hamiltonian $H_{D}$
is given by
\begin{equation}
H_{D}=\alpha _{0}\Delta +v_{_{x}}\alpha _{x}p_{x}+v_{_{y}}\alpha _{y}p_{y},
\end{equation}
where the $\alpha _{\mu }$ ($\mu =0,x,y$) matrixes satisfy the
Grassmanian algebra $\alpha _{\mu }\alpha _{\mu ^{\prime }}+\alpha
_{\mu ^{\prime }}\alpha _{\mu }=2\delta _{\mu \mu ^{\prime }}$,
and for the 2+1 dimensional system, they can be taken as the three
Pauli matrixes $\sigma _{z},\sigma _{x},\sigma _{y}$
\cite{Semenoff}.

In the above, through an analogy to the graphene physics, we have
shown how to realize massive and massless Dirac fermions with cold
atoms in an anisotropic hexagonal lattice. A more important
question for this system, however, is how to experimentally verify
the existence of these relativistic quasiparticles and the
associated  quantum phase transition. The detection method for
ultracold atoms is very different from that for condensed matter
materials, and the widely used technique based on the transport
measurements for the latter is typically not available for the
atoms. Nevertheless, there are some specific detection methods for
the trapped atomic gas, and in the following we show how to
confirm the relativistic quasiparticles and the  phase transition
with the density profile measurement \cite{ketterle,Zwierlein,16}
and the Bragg spectroscopy\cite{Ketterle99}.

The density profile of the trapped atoms can be measured through
the time-of-flight imaging with the light absorption. For free
fermions, we have ballistic expansion, and from the final measured
absorption images, one can reconstruct the initial real-space
density profile of the trapped gas \cite {16}. Now we show this
density profile provides critical information for both massive and
massless Dirac fermions. For trapped fermions, the local density
approximation (LDA) is typically well satisfied, and under the
LDA, the local chemical potential varies with the radial
coordinate by $\mu =\mu _{0}-V(\mathbf{r}),$ where $\mu _{0}$ is
the chemical potential at the trap center and
$V(\mathbf{r})=m\omega ^{2}\mathbf{r}^{2}/2$ is the global
harmonic trapping potential \cite{note1}. So, $\mu $ is a
monotonic function of $r$, and the density profile $n(r)$ is
uniquely determined by the equation of the state $n(\mu )$.

For this system at temperature $T$, the atomic density (the number
per unit cell) is given by
\begin{equation}
n(\mu )=\frac{1}{S_{0}}\int f(k_{x},k_{y},\mu )dk_{x}dk_{y},
\label{6}
\end{equation}
where $S_{0}=8\pi ^{2}/%
\sqrt{3}a^{2}$ is the area of the first Brillouin zone of the
honeycomb lattice, and $f(k_{x},k_{y},\mu )=1/\{\exp
[(E_{\mathbf{k}}-\mu )/T]+1\}$ is the Fermi distribution. At low
temperature ($T\sim 0)$, this density profile $n(\mu )$ is shown
in Fig. 2(a) and 2(b) for the parameters with massless and massive
Dirac quasiparticles, respectively. One can clearly see that for
the gapped phase with massive Dirac fermions, one has a plateau at
the atom density $n=1$ in the density profile. For the case with
massless Dirac fermions, there is no such a plateau. So the
plateau is associated with massive quasiparticles, and its
emergence with tuning of the lattice anisotropy provides an
unambiguous signal for the quantum phase transition between the
two cases.

To further confirm the massless Dirac fermions, one needs to have evidence
for their linear dispersion relation with the Dirac cone structure. At $T%
\sim 0$, $n(\mu )$ in Eq. (6) simplifies to $n(\mu
)=(1/S_{0})\int_{E_{q}\leq \mu }dk_{x}dk_{y}.$ Around the half
filling ($n=1$ which corresponds to the touching point in Fig.
1(e)), with a variation $\delta \mu $ in the chemical potential,
$\delta n(\mu )$ geometrically represents the cross section of the
Dirac cone, so it must be proportional to $\left( \delta \mu
\right) ^{2}$. In Fig. 2(c) and 2(d), we show the numerically
calculated derivative of $n(\mu )$, and indeed at the vicinity of
the half filling, we see that $\frac{\partial n}{\partial \mu }$
is linearly
proportional to $\delta \mu $ with the explicit asymptotic expression $\frac{%
\partial n}{\partial \mu }=\frac{4\pi }{v^{2}S_{0}}\left| \delta \mu \right|
$, where we have assumed the velocity $v_{x}=v_{y}=v$. So, experimentally,
from the measured density profile $n(\mu )$, one can determine its slope.
The latter quantity, with its linear form shown in Fig. 2(d), signals the
linear dispersion relation around the Dirac cone, which confirms the
massless Dirac fermions.

\begin{figure}[tbph]
\label{fig2} \includegraphics[height=5cm]{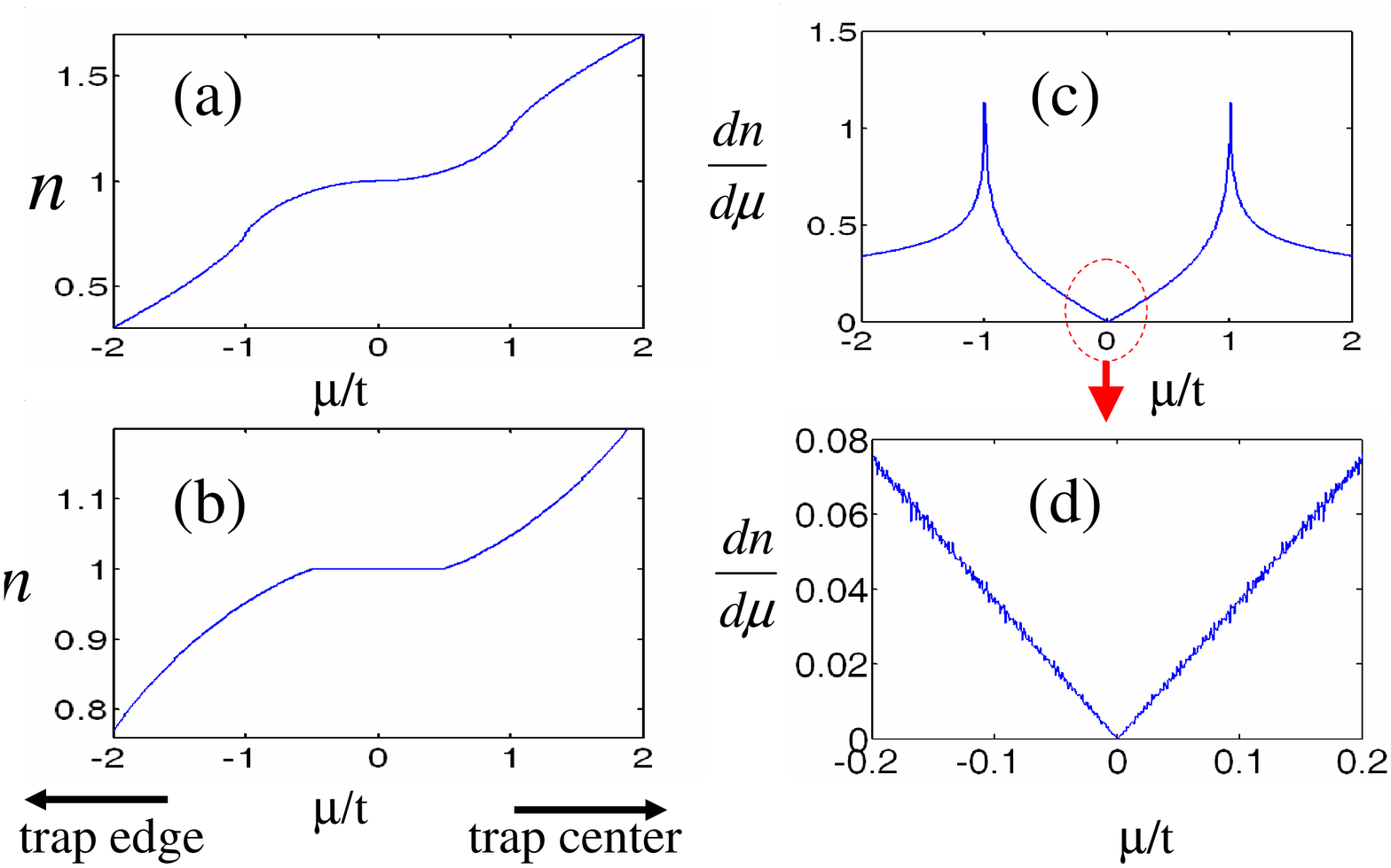} 
\caption{ (Color online) The number density of atoms $n$ per unit
cell as a function of the chemical potential $\protect\mu $
(corresponding to a rescaled atomic density profile in a trap) for
(a) $\protect\beta =1$, and (b) $\protect\beta =2.5$. A plateau
with a width $2\protect\beta -4$ appears for the latter case which
corresponds to the case when the chemical potential sweeps inside
the energy gap. (c) The derivative $dn/d\protect\mu $ as a
function of the chemical
potential $\protect\mu $ for $\protect\beta =1$. (d) An enlarged part of $%
dn/d\protect\mu $ at the vicinity of $\protect\mu =0$. The
linearity of the curve shows the linear dispersion relation for
the quasiparticles. }
\end{figure}

The Bragg spectroscopy can provide an alternative and
complementary method to confirm the linear dispersion relation for
the massless Dirac fermions and the energy gap for the massive
ones. In Bragg spectroscopy\cite{Ketterle99}, one shines two laser
beams on the atomic gas as shown in Fig. 3a. By fixing the angle
between the two beams (thus fixing the relative momentum transfer $\mathbf{%
q=k}_{2}-\mathbf{k}_{1}$, where $\mathbf{k}_{i}$ denotes the wave vector of
each laser beam), one can measure the atomic (or photonic) transition rate
by scanning the laser frequency difference $\omega =\omega _{2}-\omega _{1}$%
. From the fermi's golden rule, this transition rate basically
measures the following dynamical structure factor
\cite{Ketterle99}
\begin{equation}
S(\mathbf{q},\omega )=\sum_{\mathbf{k}_{1},\mathbf{k}_{2}}|\langle f_{%
\mathbf{k}_{2}}|H_{B}|i_{\mathbf{k}_{1}}\rangle |^{2}\delta
\lbrack \hbar \omega -E_{f\mathbf{k}_{2}}+E_{i\mathbf{k}_{1}}],
\label{D_factor}
\end{equation}
where $H_{B}=\sum_{\mathbf{k}_{1},\mathbf{k}_{2}}\Omega e^{i\mathbf{q\cdot r}%
}|i_{\mathbf{k}_{1}}\rangle \langle f_{\mathbf{k}_{2}}|+h.c.$
is the light-atom interaction Hamiltonian, and $|i_{\mathbf{k}%
_{1}}\rangle $ and $|f_{\mathbf{k}_{2}}\rangle $ denote the
initial and the final atomic states with the energies
$E_{i\mathbf{k}_{1}}$ and $E_{f\mathbf{k}_{2}}$ and the momenta
$\mathbf{k}_{1}$ and $\mathbf{k}_{2}$, respectively. At the half
filling, the valence band (the lower half of Fig. (1e) and 1(f))
is fully occupied, and the conduction band (the upper half) is
empty. In that case, the excitations are dominantly around the
touching point, and we can use the approximate dispersion relation
in Eq.(\ref{Dispersion}). For the isotropic case ($\beta =1$) with
massless Dirac Fermions, we find $S(q,\omega )$ has the expression

\begin{equation}
S(q,\omega )=\left\{
\begin{array}{ll}
0, & (\omega \leq \omega _{r}) \\
\frac{\pi \Omega ^{2}}{8v_{F}}\frac{2q_{r}^{2}-q^{2}}{\sqrt{q_{r}^{2}-q^{2}}}%
, & (\omega >\omega _{r})
\end{array}
\right.  \label{S1}
\end{equation}
where $\omega _{r}=qv/\hbar $ ($q\equiv \left| \mathbf{q}\right| $) and $%
q_{r}=\hbar \omega /v$. This dynamical structure factor is shown
in Fig. 3(b). Note that in this case, the lower cutoff frequency
$\omega _{r}$ is linearly proportional to the momentum difference
$q$, and $\omega _{r}$ vanishes when $q$ tends to zero. The ratio
between $\omega _{r}$ and $q$ gives the fermi velocity $v$, an
important parameter as the analogy of the light velocity for
conventional relativistic particles. For the anisotropic case with
$\beta >2$, the spectrum in Eq. (\ref{Dispersion}) becomes
quadratic with $E\approx \pm (\Delta+\hbar
^{2}q_{x}^{2}/2m_{x}+\hbar ^{2}q_{y}^{2}/2m_{y})$ for small
momentum transfer $\mathbf{q}$, where the effective mass
$m_{x,y}=\hbar ^{2}\Delta/v_{_{x},_{y}}^{2}$. The dynamic
structure factor in this non-relativistic limit becomes

\begin{equation}
S(q,\omega )=\left\{
\begin{array}{ll}
0, & (\omega \leq \omega _{c}^{x,y}) \\
\frac{\pi \Omega ^{2}\Delta }{2v_{_{x}}v_{_{y}}}, & (\omega >\omega
_{c}^{x,y})
\end{array}
\right.  \label{S2}
\end{equation}
where $\omega _{c}^{x,y}=2\Delta +\hbar ^{2}q_{x,y}^{2}/4m_{x,y}$.
Its form is shown in Fig. 3(b). The lower cutoff frequency $\omega
_{c}^{x,y}$ in this case does not vanish as the momentum transfer
go to zero. This distinctive difference between the dynamical
structure factors in Eqs. (\ref{S1}) and (\ref{S2}) can be used to
distinguish the cases with massive or massless Dirac fermions.
From the variation of the cutoff frequency $\omega _{c}^{x,y}$ as
a function of the momentum transfer $q_{x,y}$, one can also
experimentally figure out the important parameters such as the
energy gap $\Delta $ and the effective masses $m_{x}$ and $m_{y}$.

\begin{figure}[tbph]
\label{fig3} \includegraphics[height=3.5cm]{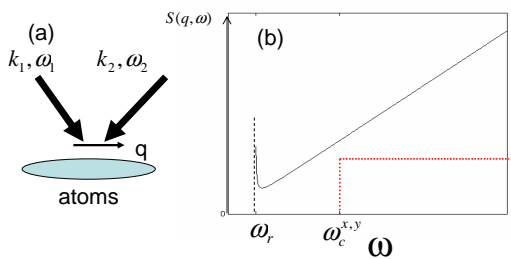} \caption{
(Color online) Schematic of the Bragg scattering. (a) Atoms are
illuminated by two laser beams with wave vectors $\mathbf{k}_{1}$, $\mathbf{k%
}_{2}$ and frequencies $\protect\omega _{1}$, $\protect\omega
_{2}$, respectively. (b) The dynamic structure factors
$S(q,\protect\omega )$ (not scaled) for the massless Dirac
fermions (solid line) and for the massive ones in the
non-relativistic limit (dotted line). The experimentally
measurable quantities $\protect%
\omega _{r}$ and $\protect\omega _{c}^{x,y}$ give important
parameters for the quasiparticles. }
\end{figure}

In summary, we have proposed an experimental scheme to simulate
and observe relativistic Dirac fermions and a  quantum phase
transition with cold atoms in a (generally anisotropic) hexagonal
optical lattice. The characteristic dispersion relations for the
massless or the massive Dirac fermions can be confirmed through
either the density profile measurement or the Bragg spectroscopy.
The  phase transition can be identified with the appearance of a
plateau in the density profile as one tunes the lattice
anisotropy. The appearance of relativistic quasiparticles,
together with control of gauge fields for this system
\cite{Osterloh,Juzeliunas}, opens up the prospect to use the
ultracold atoms to simulate some high energy physics.

This work was supported by the NSF under grant number 0431476, the
ARDA under ARO contracts, the A. P. Sloan Foundation, the State
Key Program for Basic Research of China (No. 2006CB921800), NCET
and NSFC (Nos. 10674049, 10474034, and No. 60421003).

\end{document}